\documentclass[doublecol]{epl2}

\usepackage{booktabs}
\usepackage{colortbl}
\usepackage{xcolor}
\usepackage[]{algorithm2e}
\usepackage{amsmath, amssymb}
\usepackage{mathpazo}
\usepackage{bm}
\usepackage{environ}
\usepackage{mathrsfs}
\usepackage{array,arydshln}

\usepackage{graphicx,epsfig}
\usepackage{hyperref}

\usepackage[T1]{fontenc}
\usepackage{amsmath,blkarray}

\newcommand{\be}{\begin{equation}}
\newcommand{\ee}{\end{equation}}

\title{An evolutionary game model for behavioral gambit of loyalists: \\ Global awareness and risk-aversion}
\shorttitle{An evolutionary game model for behavioral gambit of loyalists \\ Global awareness and risk-aversion} 

\author{E. Alfinito\inst{1} \and A. Barra\inst{2,3} \and M. Beccaria\inst{2,3} \and A. Fachechi\inst{2,3} \and G. Macorini\inst{2}}
\shortauthor{G. Macorini \etal}

\institute{
  \inst{1} Dipartimento di Ingegneria dell'Innovazione, Universit$\grave{a}$ del Salento, Italy\\
  \inst{2} Dipartimento di Matematica e Fisica Ennio De Giorgi, Universit$\grave{a}$ del Salento, Italy\\
  \inst{3} Istituto Nazionale di Fisica Nucleare, Sezione di Lecce, Italy
}
\pacs{89.75.Kd}{pattern formation in complex systems}
\pacs{89.75.Fb}{self organization in complex systems}
\pacs{64.60.aq}{networks and phase transitions}

\abstract{We study the phase diagram of a minority game where three classes of agents are present.
Two types of agents play a risk-loving game that we model by the standard Snowdrift Game.
The behaviour of the third type of agents is coded by {\em indifference} w.r.t. the game at all: their dynamics is designed to account for risk-aversion as an innovative behavioral gambit. From this point of view, the choice of this solitary strategy
is enhanced when innovation starts, while is depressed when it becomes the majority option.
This implies that the payoff matrix of the game becomes dependent on the global awareness
of the agents measured by the relevance of the population of the indifferent players. The resulting dynamics is
non-trivial with different kinds of phase transition depending on a few model parameters. The
phase diagram is studied on regular as well as complex networks.}

\begin{document}

\maketitle

\section{Introduction}

Risk and uncertainty are important elements of human behavior, in particular when dealing with
interactions of international teams and global counterparts. Here, by risk we just mean the intrinsic uncertainty in a multi-strategy individual interaction/game characterized by a pay-off depending non-trivially on the mutual strategic choices. The level of risk or, generally speaking, uncertainty that may be tolerable is not constant over different cultures, see for instance
 \cite{hofstede2001culture,hofstede2010cultures}. It appears that risk aversion attitude is correlated with a low amount
of tolerance for vague or ambiguous behaviours. In such a case, structured sets of rules are preferred over more undetermined
situations. This implies a sort of cultural payoff component associated with playing risk-averse strategies.
Besides, cultural conformity is  a key factor in the evolution of complex culture in humans
\cite{rendell2011cognitive,van2013potent,whiten2011culture} as well as in
animals \cite{aplin2015experimentally}. On the other hand,
there is tension with the drive to innovate. Innovation is an important ingredient to promote longstanding  high levels of welfare, as it happens in real markets \cite{balkin2000ceo,lyon2002enhancing,wolfe1994organizational}. Deviation from conformity
is an important evolutive mechanism, the so-called behavioral gambit, see for instance
\cite{fawcett2013exposing}. In  this paper we propose a simple model describing how innovation
is triggered when a population component is risk-averse in presence of many individuals accepting
playing some risky competition requiring cooperation. We shall assume a tendency to
play risk averse strategy with enhanced payoff for the above mentioned cultural reasons. On the
other hand, we shall also assume that an increased sense of safety due to a large number of
risk-averse players will feedback the strategy choice and reduce the extra payoff allowing for more
risk-loving attitudes.
In the context of epidemics disease critical spreading \cite{our-sci-rep},
it is also natural to assume that the effective infection
probability is affected by the perception of the risk of being infected. This has been
assumed to be  related to the fraction of infected neighbors in the recent models
discussed in \cite{bagnoli2007risk,massaro2014epidemic,our-epl}.
The precise mechanisms for this to happen are an important issue, like
aggregation effects  \cite{chambers2012does} or the role of the endorsers \cite{jiuan1999strategies}.
Here, we shall explore the dynamical balance between  risk-aversion and
the opposite  behaviour induced by global awareness of the fact that such attitude
is the majority option: a natural application of the present model deals with the phenomenon of {\it nonconformism}, a tendency ranging from social and political settings to economic or financial contexts. Nonconformist individuals refuse to take part in the game, and their behavior does not depend on the interactions with other competitors. They also tend to arrange in closed subcommunities sharing the same behavior (or opinion). A distinctive feature of the phenomenon is that the nonconformist option is discouraged if this behavior becomes a mainstream mindset.
\newline
We shall describe this competition by means of  evolutionary game theory \cite{smith1982evolution,weibull1997evolutionary,nowak2006evolutionary,sigmund2010calculus}.
This is  a theoretical framework where
emergence of collective behaviour may be observed,  especially when played on structured populations
\cite{szabo2007evolutionary,roca2009evolutionary}. We shall adopt the {Snowdrift Game} as our background model of competition \cite{doebeli2005models}.
 In the Snowdrift Game \cite{santos2012dynamics} cooperating players can coexist with defector ones
 even in well-mixed systems. The spatial structure is important and
 may even hinder the evolution of  cooperation \cite{hauert2004spatial}.
More recently, Santos and Pacheco  \cite{santos2005scale}
  discovered that cooperation generically emerges on scale-free networks. In addition, Szab\'{o} et al.  \cite{szabo1998evolutionary,szabo2002phase,szabo2004cooperation}
    presented a stochastic evolutionary rule aimed to capture the bounded rationality feature in order to
characterize real system game dynamics.
\newline
We focus on the memory-based Snowdrift Game  proposed in \cite{wang2006memory}. In this model, the effects of the individuals memory is taken into account in the determination of  the evolution of strategies: players update their strategies based on their past experience.  This approach is based on \cite{challet1997emergence}, where a so-called {\em minority game} \cite{Coolen,Javarone} is considered  in which agents
 make decisions exclusively according to the common information stored in their memories. It is found that finite memories of agents have crucial effects on the dynamics of the minority game  \cite{challet2008dynamical,jefferies2001market}.
\newline
To take into account the risk-averse strategy option we included {\em solitary players} \cite{zhong2006evolutionary,zhong2007evolutionary}. In the standard snowdrift game with these loners, one has a solitary-strategy describing people choosing not to participate in the competition and would prefer to take a fixed albeit small  payoff. Hauert et al. studied the effects of their presence in a generalization of the prisoner dilemma game called the {\em public goods game} \cite{hauert2002volunteering,szabo2002phase}.
This basic model will be modified in order to take into account two novel mechanisms according to the previous discussion. The first is a cultural payoff enhancements for those players choosing to avoid the snowdrift game strategies. The second is a damping effect of the risk-aversion tendency. It depends on the global awareness of the system, {\em i.e.} on the fraction of players preferring the risk-averse strategy, in the spirit of
\cite{bagnoli2007risk}.

\section{The Modified Snowdrift Game}
In \cite{li-xin} the authors resumed  the SG model with the loners in the evolutionary setting, while memory in the system has been considered for the pure SG in \cite{wang}. The main results of these two papers will be  summarized in the Supplementary Material.
\newline
Our modification of the Snowdrift game with S players is based on the payoff matrix
\be\small
\mathbf{P} =
\begin{blockarray}{cccc}
& \mathbf{C} & \mathbf{D} & \mathbf{S}  \\
\begin{block}{c(ccc)}
\mathbf{C} & 1 & 1-r & q  \\
\mathbf{D} & 1+r & 0 & q   \\
\mathbf{S} & (1+\xi)\,q & (1+\xi)\,q&  (1+\xi)\,q \\
\end{block}
\end{blockarray},
 \ee
where $r$ is the cost-to-reward ratio, $q$ a fixed constant and at each round
\be\small
\xi = 1- f_{S},
\ee
$f_S$ being the fraction of players in the $S$-strategy. The payoff for pair of strategies involving solitaries is standard for $\xi=0$. Adopting the solitary strategy has an
additional payoff $\xi\,q$ decreasing when the number of $S$-players increases. The rules of this
memory-based evolutionary game are the following: let us consider a generic graph; each player lies on a node of this graph such that pairwise connected players challenge reciprocally match by match and this happens for all the (connected) couples. The total payoff of any player is simply the linear sum of the payoff collected in each single duel. Once a round is terminated, all the players evaluate their performances  by playing a virtual match where they use the anti-strategy w.r.t. the one they've just adopted in order to calculate their potential reward with this revised settings: if the latter is actually better then the employed one (i.e., if the player obtained higher score with the anti-strategy), it keeps the latter as the best strategy to be used and it stores this information in its memory (whose length is fixed to $M$ bits\footnote{We deepen this point in the Supplementary Material.}). Once a new game starts, the probability of making a decision ({\it i.e.}, choosing C, D or S) depends on the relative ratio of the numbers of C, D and S stored in the player's memories, then, they all update their memories simultaneously and, by iterating the outlines steps, the game's dynamics takes place.
\newline
Note that the parameter $\xi$ introduces a global feedback in the score of the S players, which
now depends not only on the local interaction but also on the global state of the community.
\newline
We introduce, as order parameters, the three fractions of players
$$\small
f_C \equiv \frac{C}{N}, \ \ \ f_D \equiv \frac{D}{N}, \ \ \ f_S \equiv \frac{S}{N},
$$
(where $N=C+D+S$), and we will study them as functions of $r$ for fixed values of $q = \{0.1,\dots, 0.4\}$ and viceversa \footnote{Numerically, the curves are obtained averaging over 100 Monte Carlo simulations for any value of $r$ and $q$.}.
\newline
Note further that a pure SG with memory-driven evolution has been considered in \cite{wang}, both in regular lattices and complex networks, where it has been shown that taking the memory Markovian and with finite length, the evolution of the system  always reaches an (almost) stable state,
characterized by a (roughly) constant frequency of cooperators.
In the regular lattice, cooperators and defectors arrange in
typical spatial patterns and  $f_C$ has a {\em step-shape} as function of  $r$,
the number of step being dictated by the number of neighbors.
\begin{figure}[htb]
  \centering
    \includegraphics[width=0.8\columnwidth]{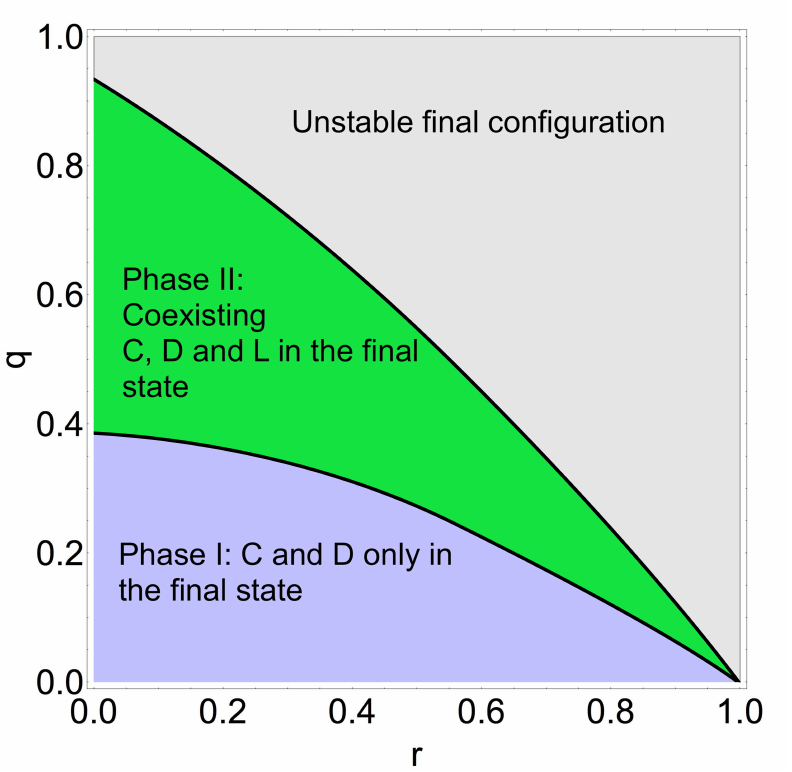}
      \caption{\label{fig3a} Phase diagram of the model.  The black lines mark the transitions
      from the first phase without solitaries to the second with the three strategies coexisting and from the second to the (unstable) third one.}
      \end{figure}
A variation of the SG theme with the S-players has been instead analyzed in \cite{li-xin}, where the authors focused on completely connected graphs and regular (with four neighbors) lattices. Their main results are that, in a completely connected graph, the
coexistence of the three characters is impossible: C and D players can exist only in the absence of solitary category, or the latter
take over the entire population. For the regular graph, instead, the presence of the S-players leads to an
improvement of the collaborative attitude, coexistence is possible, and in general the C-players can  survive in the
full parameter space of the model.

\section{Regular lattice $R_8$}

In this section we discuss the results for our extension in a regular lattice with 8 neighbors.
The phase diagram of the model can be divided in three main regions in the $(r,q) = [0,1] \times [0,1]$ square,
as shown in the phase diagram presented in figure \ref{fig3a}. In the first and second phases the system evolves to a steady state with
coexisting C-D and C-D-S players respectively. In the third phase, the players never reach  a stable strategy choice.
In the first phase (highlighted in blue), for $r$ and $q$ small enough, the  system evolves toward a stable final state without S-players.
The final state is the same as it would be for a community without the  S-players at all, and the step-like shapes of the $f_{C,D}$ fractions
are evident from the plots of the densities  vs $r$ as reported in figure \ref{fig4a}.
In the second phase (highlighted in green), the system evolves almost everywhere (see below) to a stable final configuration characterized by the coexistence of the three kinds of players. However  the total fraction of S individuals, as the size of their clusters, increase both with $q$ and $r$, as a glance at both figures \ref{fig4a}, \ref{fig4b} may confirm.
\begin{figure}[htb]
  \centering
     \includegraphics[width=0.98\columnwidth]{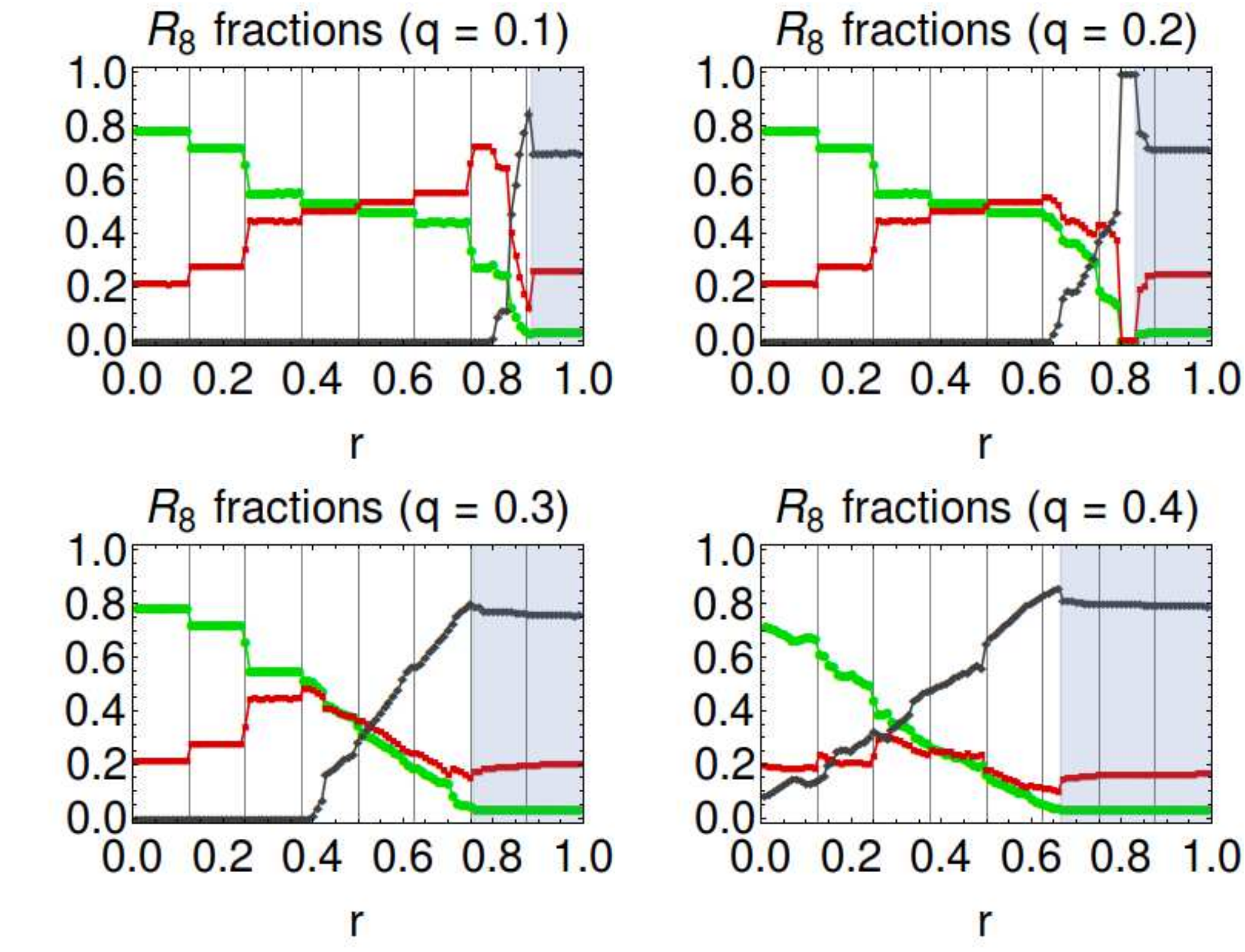}
     \caption{\label{fig4a}
      Fractions of cooperators $f_C$, defectors $f_D$ and solitary players $f_S$ as functions of the parameter $r$ for $q=0.1, \dots 0.4$. 
      Green dots denotes $f_C$, red $f_D$ and gray $f_S$.
      The shaded regions in the plots correspond to the third phase of the system where no stable spatial arrangement is reached. Gray vertical lines correspond to
      the phase transitions in the SG without S-players. Further analysis is presented in the Supplementary Material.}
\end{figure}

\begin{figure}[htb]
  \centering
     \includegraphics[width=0.98\columnwidth]{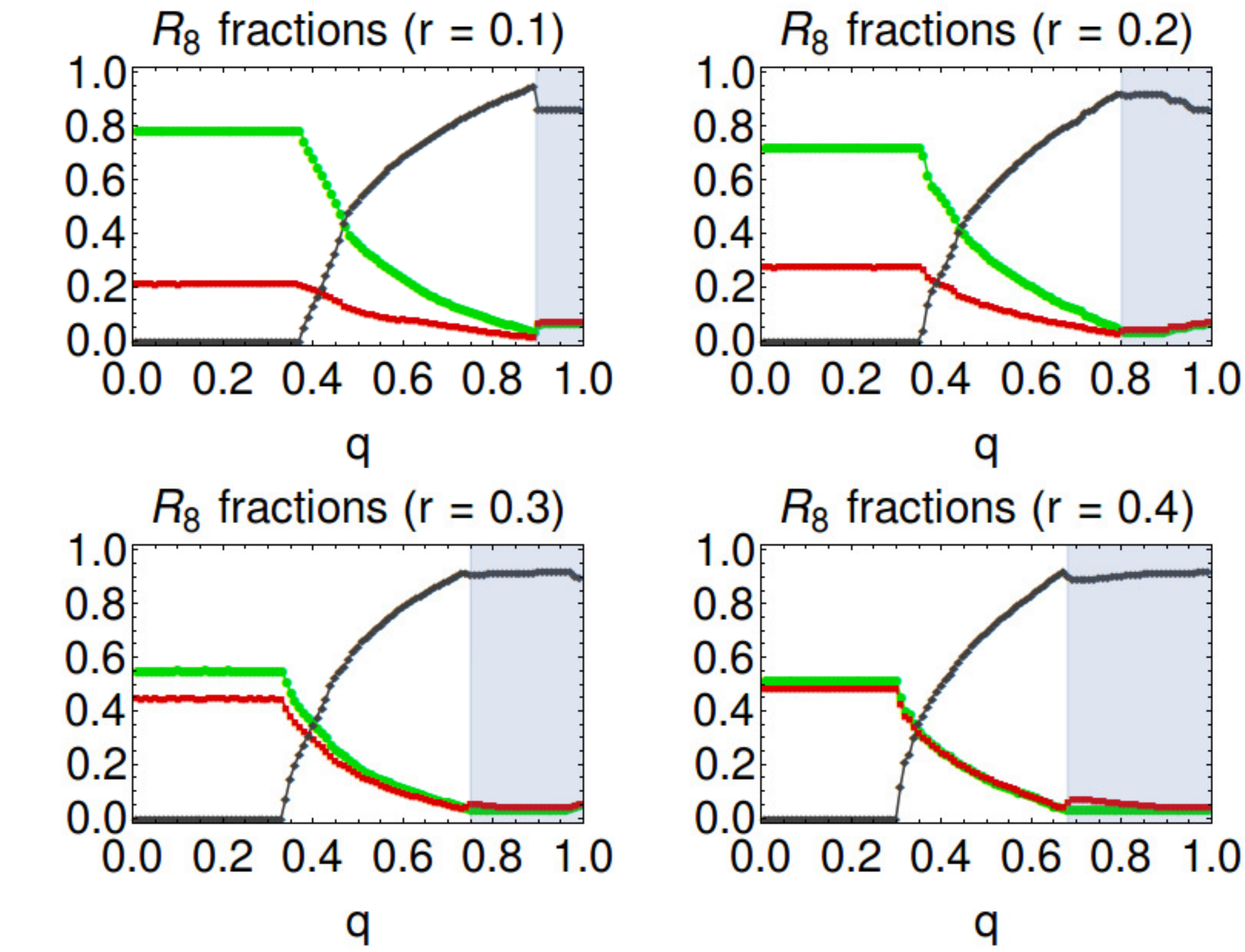}
     \caption{\label{fig4b}
       Fractions of cooperators $f_C$, defectors $f_D$ and solitary players $f_S$ as functions of the parameter $q$ for $r=0.1, \dots 0.4$. 
      Green dots denotes $f_C$, red $f_D$ and gray dots $f_S$.
      The shaded regions in the plots correspond to the third phase of the system where no stable spatial arrangement is reached.}
\end{figure}
This phase is characterized by typical spatial configurations, where S-players arrange forming clustered sub-communities.
We stress that this stable spatial arrangement of the strategies is strikingly different from the results of \cite{li-xin}.
Both the stability of the final state and the regularity of the patterns are
consequences of the memory-based update for the strategies of the players.
Despite the phase diagram of the model of \cite{li-xin}  is qualitatively similar to the present one, the probabilistic update of their players
 forbids the stability of the choices and consequently the formation of the patterns.
The rearrangement in sub-communities by the S-players is instead dictated by the nonlinear feedback of $\xi$ in the payoff matrix, favouring the grouping of players that prefer not to participate.
\newline
Looking at the various $q=0.1, 0.2, ...$ in figure \ref{fig4a},  one can clearly see that once the solitaries appear, their fraction becomes quickly dominant increasing $r$: a higher payoff for the defectors means a higher risk to be cheated for the C-players, and progressively more and more
individuals find advantageous to avoid this risk, choosing for the solitary behavior.
\newline
The fraction $f_S$ is an increasing function of both $r$ and $q$  but in rather different ways: while the transitions of $f_S$ in $r$ are first-order (i.e. there is a genuine {\em jump} by the order parameter at the critical values $r_c$, see figure \ref{fig4a})\footnote{We discuss in the Supplementary Material how to locate these critical $r_C$ values by a standard stability analysis.}, the transition from zero to non-zero values for $f_S$ vs $q$ resembles a critical (second-order) scenario: the order parameter continuously raises from zero at $q_C$, see figure \ref{fig4b} (further there is a genuine divergence in the susceptibility -the variance of the order parameters, suitably amplified by a volume factor- for $q \to q_C$, as deepened in the Supplementary Material).
\begin{figure}[htb]
  \centering
    \includegraphics[width=0.98\columnwidth]{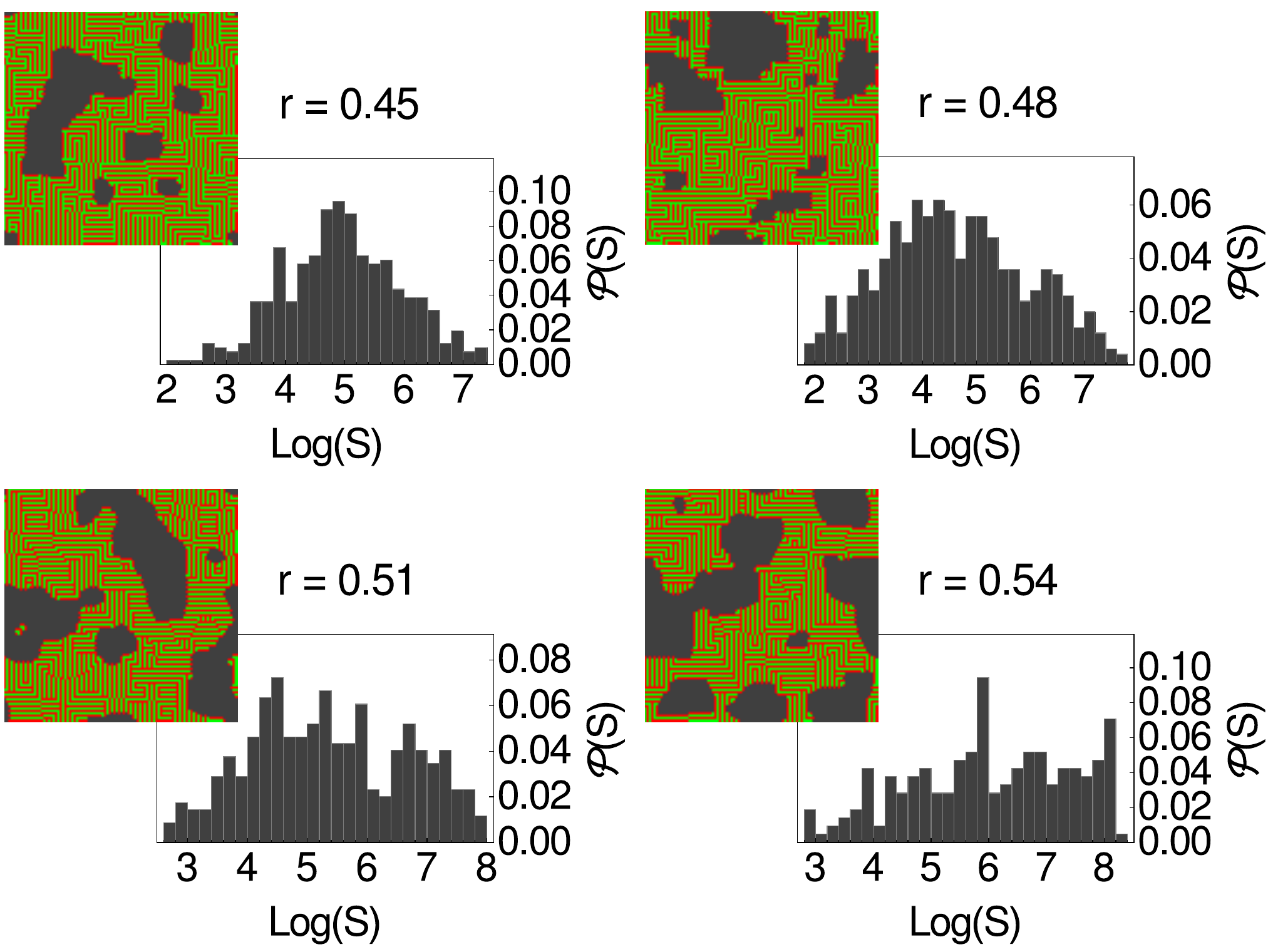}
      \caption{\label{snaps}
Probability distributions (and relative snapshots) for the size $S$ of S-clusters (in gray), in a $100 \times 100$ lattice for fixed $q=0.3$
and varying $r$ around the percolation threshold for a giant S-component.}
\end{figure}
We aim to better characterize phase II (since understanding network's dynamics in that phase is crucial to see why there is a third, unstable, phase in the phase diagram) by looking at the evolution of the size distribution of the S-clusters: in figure \ref{snaps} we show the probability distributions for the size of S-clusters, in a $100 \times 100$ lattice at fixed $q=0.3$ and varying $r$ around the percolation threshold for a S-players' giant component. These loners appears in a relatively large number of small islands progressively increasing their size and eroding the regular pattern of C- and D-players, eventually merging. As anticipated above, in this phase the system does not always reach a perfectly stable final configuration: this is because a (relatively small) fraction of players (typically located at the borders of the S-clusters) can exhibit an oscillatory  behavior. These local instabilities are closely related to an oscillating balance in the payoff matrix among the possible gains, due to the nonlinear feedback of $\xi$: this results in a perpetual indecisiveness for some players as a structural property for models with global awareness.
\newline
The growth of the S-density leads eventually to another transition\footnote{The nature of the transition is deepened in the Supplementary Material.} toward the third phase, where the system is intrinsically unstable: all the players are never able to make a definitive choice for a strategy.
Indeed, for large enough $q$, the incentive not to participate to the SG would make the S-strategy the preferred choice for all the individuals and
the system would evolve toward a state where the S-players completely dominate the system.
Nevertheless, a configuration with only S players is not  stable. With $f_S = 1$ and $\xi=0$ in the payoff matrix,
all the players would  get exactly the same payoff for any  strategy they choose (resulting in a random update rule),
leading to a destabilization of the system configuration. Finally, we note that, despite a spatial arrangement is impossible,
yet the model is still characterized by some constant (averaged in time) fractions for the strategies of the players in this third phase, as can be clearly seen in the gray regions of the plots in figures \ref{fig4a},\ \ref{fig4b}, but the collaborative attitude is highly inhibited.

\section{Complex Networks}

The SG with S-players can be extended as well on complex networks. In this section, we consider the model on two typical examples of complex networks, namely Watts-Strogatz (WS) \cite{wsmodel} and Barab\`asi-Albert (BA) \cite{bamodel} graphs.

\subsection{Watts-Strogatz Networks}

\begin{figure}[htb]
  \centering
    \includegraphics[width=0.98\columnwidth]{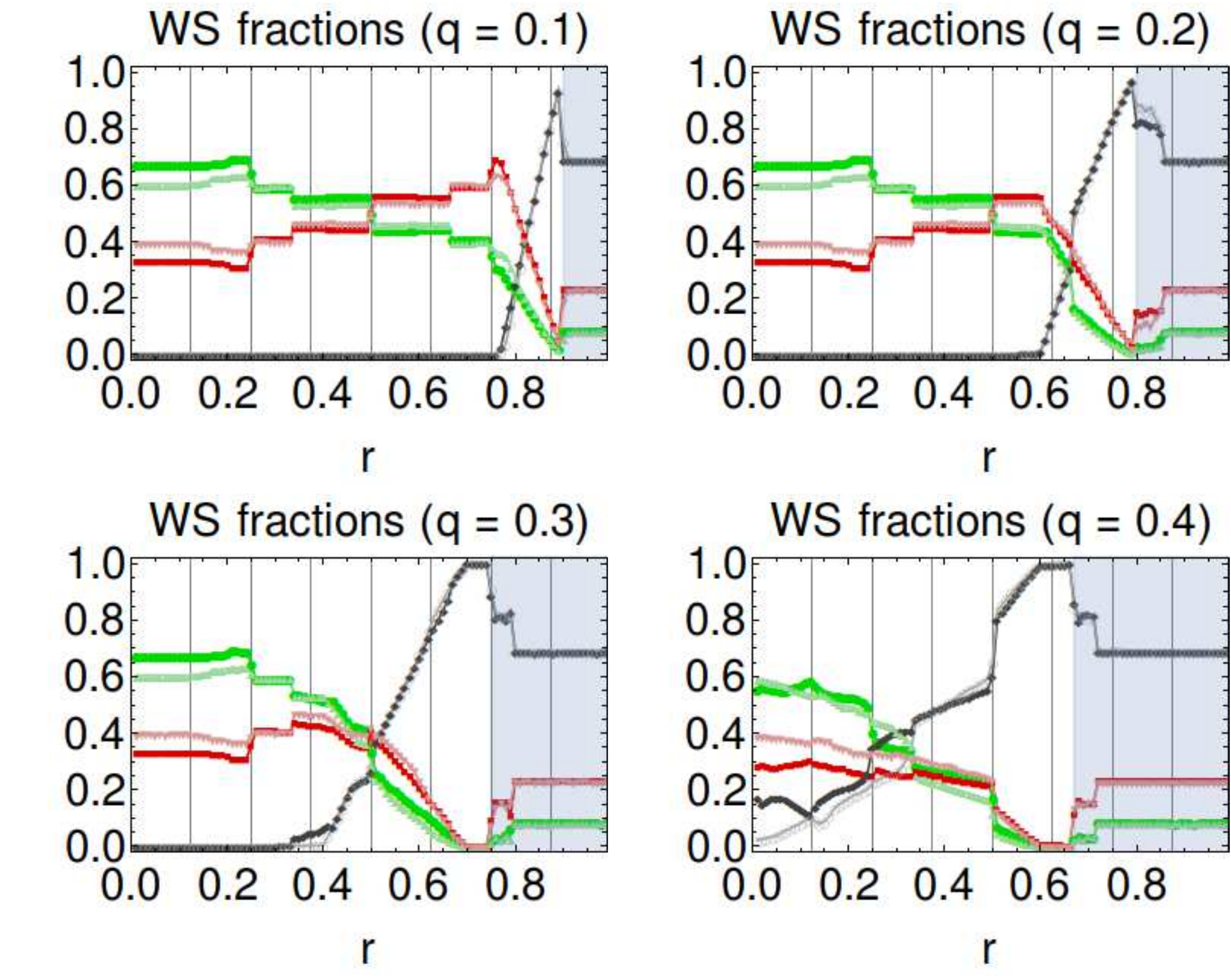}
          \caption{\label{figws1}Fractions of cooperators $f_C$, defectors $f_D$ and solitary players  $f_S$ as functions of the parameter $r$, for a WS network with $\theta = 0.1$ (darker colors)
          and $\theta = 0.5$ (lighter colors). Green dots denote $f_C$, red $f_D$ and gray $f_S$.
          The shaded regions in the plots correspond to the third phase of the system, where no stable spatial arrangement is reached.
           }
\end{figure}

In the Watts-Strogatz case, we consider two realizations of the networks for different
rewiring probabilities ($\theta = 0.1$ and $\theta = 0.5$) in the algorithmic construction of the topology.
The global picture for the relationships among the characters resembles
the results obtained in the regular graph and the phase space is qualitatively very similar.
The evolution of the fractions of C, D and S players is summarized
in figure \ref{figws1} as a function of the parameter $r$ in the payoff matrix (for fixed $q$ values).
%

As in the regular case, low values of $r$ lead to the extinction of solitary players.
Increasing $r$, the S-density in the final configuration is non-vanishing and takes over quickly the majority of the nodes  (again,
when the fraction of the these players approaches 1 the final configuration becomes unstable). Looking at the fractions  for $\theta = 0.1$ and
$\theta = 0.5$, one can immediately see that the results look extremely similar and qualitatively independent on the rewiring probability.
\newline
One may note that, at least in the first phase (without S-players in the final state), one can spot a reminiscence of a steps-like pattern for the C- and D-densities: the WS model is in fact realized starting from a regular graph with a rewiring prescription governed by $\theta$. This procedure preserves the mean degree, and the final degree distribution falls exponentially (in the volume) for large deviations from the mean value.
At least for small $\theta$ values, this makes the system's dynamics in the WS setting rather close to that on a regular network with the same mean degree for the nodes (essentially assuming $P(k) \to \delta(k- \bar{k})$). Then, by noticing that the evolution of the system remains qualitatively the same even for larger $\theta$ values, one could say that the SG game (with or without S-players) is rather insensitive to the {\em small-world} property of a network (since the strategy choices are almost completely dictated by local interactions among individuals).
\newline
Increasing the rewiring probability, {\it i.e.} the randomization of the final graph, has the only effect to slightly and uniformly reduce the convenience of a collaborative behavior.
\subsection{Barab\`asi-Albert Networks}
\begin{figure}[htb!]
  \centering
   \includegraphics[width=0.98\columnwidth]{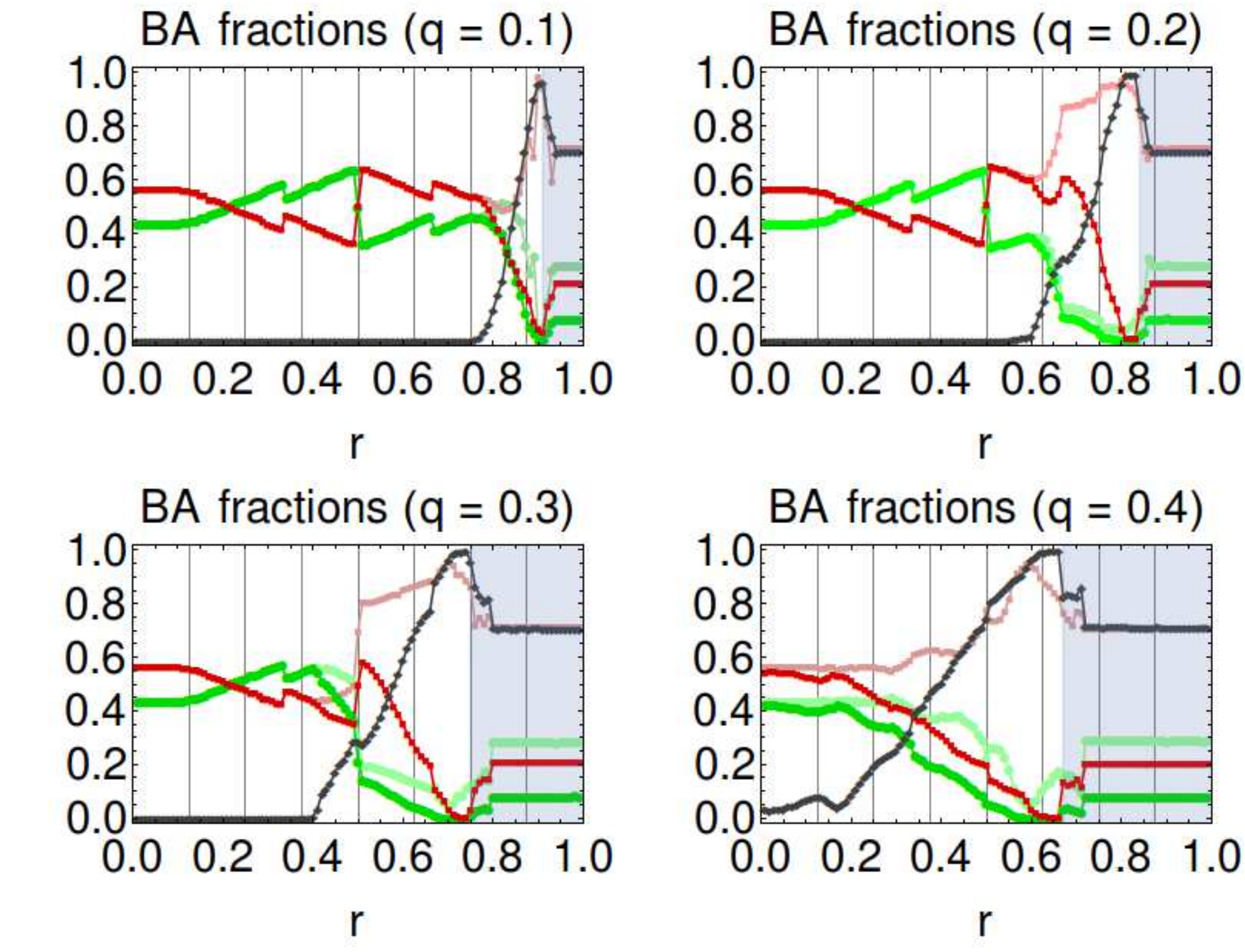}
          \caption{\label{figBA1} Fractions of cooperators $f_C$, defectors $f_D$ and solitary players  $f_S$ as functions of the parameter $r$ for different choices of $q$.
          Green dots denotes $f_C$, red $f_D$ and gray $f_S$.
          The shaded regions in the plots correspond to the third phase of the system where no stable spatial arrangement is reached.
          }
\end{figure}
\begin{figure}[htb]
  \centering
  \includegraphics[width=0.98\columnwidth]{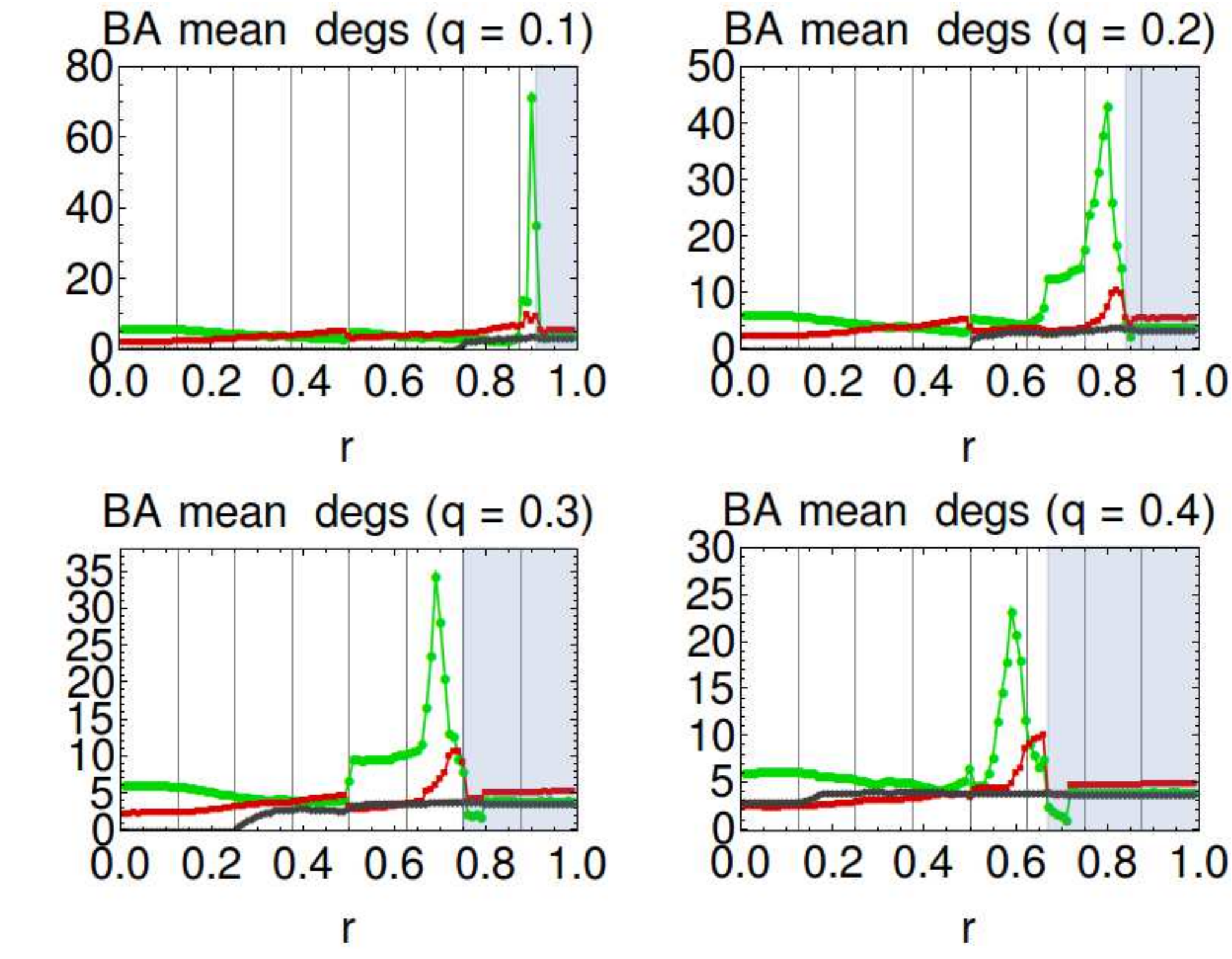}
          \caption{\label{figBA-deg} Mean degree of the C, D and S nodes respectively. Green dots denotes the mean degree of the nodes occupied by C players,
          red dots are used for D players and finally gray dots denote S players. The plots show the mean degree as function of the parameter $r$ for different choices of $q$.
          The shaded regions in the plots correspond to the third phase of the system where no stable spatial arrangement is observed.
          }
\end{figure}
The preferential attachment algorithm for the construction of the BA graphs results in a (approximatively for finite size graphs) power law distribution
of the nodes.  In this case, the local structure of the network is extremely heterogeneous.
From the results of the previous sections, one can thus expect that the SG  with S-players is characterized by a totally different evolution in the BA setting.
This expectation is confirmed by \cite{wang}, where the authors considered the pure SG with memory in scale-free networks.
Without the  S-players, the authors  \cite{wang} found that the cooperative density has a strong, non-monotonous dependence on the parameter $r$. In particular, $f_C$ presents peaks at specific values of $r$, resulting in the non intuitive picture that a larger
payoff for a selfish behavior may lead to an enhancement of the collaboration.
As the S-players are included in the game, the above results change drastically: figure \ref{figBA1} reports the fractions of cooperators $f_C$, defectors $f_D$ and solitaries $f_S$ as functions of the parameter $r$, for different choices of $q$ in $BA_2$ graphs (we denote with $BA_2$ the Barab\`asi Albert networks constructed adding a node with two new links at any step of the growing algorithm).
Increasing $r$ for a given $q$, the  S-players take again quickly the majority of the nodes, but  their presence
 has a strong destructive effect on the collaborative behavior of the other players: the fraction of the collaborators never grows with $r$ in presence of the S players.
This is even more evident looking at the relative fractions of the C and D behaviors only, excluding the S-nodes in the normalizations, {\it i.e} $\bar{f}_{C,D} = \frac{n_{C,D}}{n_C+ n_D}$ (reported in lighter colors in figure \ref{figBA1}). The presence of the S-players strongly inhibits the collaboration,  the relative fraction $\bar f_C$ being constantly lower than $\bar f_{D}$ as the great majority of the players that do not opt for the S-strategy  are forced to be defectors.
We can better understand the evolution of the dynamics on these networks
looking at the mean degree of the nodes occupied by the various players, as shown in figure \ref{figBA-deg}. For all the values of $q$, the mean degree of the collaborating nodes has a sharp peak when $r$ grows enough such that the S-players are close to take over the network ({\it i.e.} at the onset of the transition to the unstable phase).
These plots suggest  that, while the solitaries become progressively the largest fraction increasing the payoff for the defection,
the collaborative behavior resists only in  few nodes with a relatively large degree,
where the minority of C-players is segregated and surrounded. The role of high degree nodes as bastions of least resistance for the cooperative choice is rather unusual. On the other hand, the mean degree of the S-players stays almost
constant after their appearance, suggesting that  this strategy spreads among the various players as $r$ increases in a rather uniform way.

\section{Conclusions}

By including also {\em solitary players}, in this paper, we extended the Snowdrift Game with memory to account for a third behavior, whose decisional process relies on global awareness (and not just local knowledge) in a {\em minority-game attitude}: we studied numerically the phase diagram of this model, namely the evolution of the relative fractions of the three types of players versus the tunable parameters $r$ (the standard cost-to-reward ratio) and $q$ (the solitary payoff).  We found robust numerical evidences that, starting by $r=0$ and $q=0$, along the $r$-axes, the evolution of these fractions happens trough discontinuous (first-order like) phase transitions, while its growth along the $q$-axes is smooth and coupled to a peak of their  fluctuations, as typical in critical (second-order like) phase transitions. Nevertheless, the model has a third, intrinsically unstable, phase -driven by the global awareness introduced in the payoff matrix- that is reached for high values of $(r,q)$ in a way quite similar to percolation in random graphs.
\newline
Concerning the role of the solitary strategy, we found that individuals choosing this behavior show a strong predisposition to group together in clusters for relatively high q, therefore mimicing the tendency of nonconformist people to join together in subcommunities with the same mindset (e.g. Facebook pages, blogs, organizations etc.). However, increasing further the loners' payoff q leads to the unstable phase, in which the gain in the nonconformist behavior is so high that it quickly becomes the majority option, thus turning into a conventional attitude. Then, all of the three strategies becomes equally profitable, resulting in a downsizing of the nonconformist front and making the evolution cyclic.
As future outlooks we aim to bypass the pairwise decision rule enlarging the outlined scheme to include also mixed populations \cite{attila}.

\bigskip

\bigskip

\textbf{SUPPLEMENTARY MATERIAL}

\bigskip

\bigskip

\section{The snow-drift game with memory}
\label{sec:rev}

In order to frame correctly our work, we use this first section of the Supplementary Material to briefly resume the relevant results recently appeared in the Literature.
\newline
The standard snowdrift model may be described in terms of the dilemma of
two players blocked by a snowdrift.
Each  player can choose a collaborative (C) behavior (and help to shovel, sharing the effort)
or a defective (D) attitude (in the hope to exploit the other player's effort).
Collaboration leads to a gain, but has a cost. The cost is shared if both player choose the C strategy,
but is borne by the collaborative player only if its partner choose to be defective.
While both players would maximize the personal gain
choosing a defective attitude if the other player is collaborative, they risk to gain nothing if both  opt for a defective strategy.
The two player model can be completely characterized by the following payoff matrix for
the possible combinations of C-D choices.
\be\small
  \mathbf{P}_{CD} =
    \begin{blockarray}{ccc}
        &  \mathbf{C} & \mathbf{D}  \\
      \begin{block}{c(cc)}
        \mathbf{C}  &  1 & 1-r  \\
        \mathbf{D}  &  1+r & 0   \\
      \end{block}
    \end{blockarray}
\ee
where the parameter $0 < r < 1$ represents the cost-to-reward ratio.
The model can be easily extended to a community,  where any of the individual can choose the
collaborative or defective behavior when playing
with a number of other members. The total payoff of each player is the sum over all its encounters.
In an evolutionary setting, the game is repeated over time, with the players
updating their strategies to maximize their gain.
The update of the strategy  may involve a comparison with the gains
obtained by other players or be based on the experience
{\it i.e. the memory} of the player itself.\\
A pure SG with memory-driven evolution has been considered in \cite{wang}, both in regular lattices and complex networks.
Here we briefly summarize their results for the regular lattice particularly relevant for our study.
Taking the memory length $M \geq 2$ and finite, the evolution of the system   always reaches a (almost) stable state,
characterized by a (roughly) constant density of cooperators $f_C$. 
In the regular lattice cooperators and defectors arrange in
typical spatial patterns and  $f_C$ has a {\em steps-shape} as function of  $r$,
the number of steps being dictated by the number of neighbors. This behavior of $f_C$ can be  explained looking at the local stability of
the possible arrangements of the C and D strategies on the lattice. Indeed, for the eight neighbor case,
looking at a $3 \times 3$ local square on the lattice, a central
C individual with $k$ C-neighbors and $8-k$ D-neighbors
would get a score $S_C(r) = k + (8 - k)(1- r)$. Turning to a D behavior its score would be $S_D(r) = k(1 +r)$. Comparing the two score,
the local stability imposes
\begin{equation}
S_C(r_c)=S_D(r_c),
\end{equation}
giving 8 solutions for $r_c = 1/8, 2/8, \dots,1$, which are the critical values of $r$ that mark the transitions (see figure two in the main text).
In each of these eight regions $[0 < r < 1/8], [1/8 < r < 2/8], $ $\dots$ the final density of the cooperators is constant in the absorbing state,
resulting in the steps shape of $f_C$ mentioned above.
\par\bigskip

\section{Accounting for behavioral gambit of loyalists, a summary of results}
\label{sec:model}

The presence of a third character has been studied
starting from \cite{loners1, loners2}, in the so-called Public Goods Game. The new strategy (to which we refer as "solitary" (S) or "loner") accounts for individuals choosing the risk-averse behavior. A variation for the SG game including the S-players has been already analyzed in \cite{medie}. They refuse to partecipate to the SG, obtaining as a reward a fixed income (independent of the cost-to-reward ratio $r$). The associated payoff matrix is then
\be\small
\mathbf{P}_{CDS} =
\begin{blockarray}{cccc}
& \mathbf{C} & \mathbf{D} & \mathbf{S}  \\
\begin{block}{c(ccc)}
\mathbf{C} & 1 & 1-r & q  \\
\mathbf{D} & 1+r & 0 & q   \\
\mathbf{S} & q &  q&  q \\
\end{block}
\end{blockarray}\ ,
 \ee
where $q$ is a fixed constant. In their work  \cite{medie}, the authors focused on completely connected graphs and regular (with four neighbors) lattices.

In completely connected graphs, the coexistence of the three characters is impossible: indeed, C- and D-players can only exist in absence of S-individuals, otherwise the latter take over the entire population. This result can be understood by noticing that the total payoffs only depend on the densities at the time under consideration, with each type of player having the same payoff.

On the other hand, for regular graphs the presence of solitary players leads to an improvement of the collaborative attitude, so coexistence is possible and generally C-players survive in the square $[0 < r < 1]\times[0 < q < 1]$.
Moreover, exploring the full parameter space one finds three possible phases: the first one exhibiting a uniform C-player population, the second one with coexisting C- and D-individuals, and finally the third phase with all of the three characters present. In the regular lattice, the survival of cooperators is facilitated because of the S-players (as they enhance their presence average payoff). The reason behind this peculiar character lies in the spatial restrictions imposed by the underlying topology, meaning that the individual payoff does not only depend on the character but also on the local environment the player is competing in. This is proved by the tendency of the players to cluster or aggregate of players into configurations where the payoffs are favored and the dependence of the relative equilibrium densities in $r$ reflects the change in preferred configurations as this parameter is tuned.

When restricted to its original 2 components (C and D), the memory-based snowdrift model has a peculiar behaviour.
Formation of spatial patterns are observed on lattices, together with the step structure of the density of cooperation versus the payoff $r$ parameter. The memory length of individuals plays different roles at each cooperation level.
On heterogeneous ({\it e.g.} scale-free) networks, nodes with highest degree are taken over by defectors and this leads to an increased cooperation level. Furthermore, similarly to the cases on lattices, the average degrees of SF networks is still a significant structural property for determining cooperative behavior. The memory effect on cooperation investigated in our work may draw some attention from scientific communities in the study of evolutionary games.

\section{The backbone of the model and its dynamics}

In this section, we explicitly consider the presence of the third $S$ character in the community, with two important modifications with respect to \cite{medie}.
The first one involves global feedback mechanics for the payoff of S-players, which now depends on the total S-density in the community.
The second modification is that the time-evolution is memory-based, following the proposal of \cite{wang}.
\newline
Hereafter, we comment in more details our generalization and the numerical findings of our investigation. To make the present discussion self-contained, we also re-discuss here some details already reported in the main text. In our model, the adoption of the solitary strategy has an
additional payoff $\xi\,q$ decreasing when the number of S-players increases (see the discussion below). The rules are the following: once players have been placed on the nodes of the network,
at every time step all pairs of connected players play the game simultaneously. The total payoff of each player is computed by summing over the 2-player games played with its neighbours. At the end of a  round, each player chooses its best strategy by considering the would-be payoff, at fixed
strategy choice of the other players. Finally, this optimal strategy is stored in the last bit of player's memory. Taking into account the bounded rationality of players, we assume that players are limited in their analyzing power by allowing them to retain at most the last $M$ bits of past strategies information. In the next iteration, the probability of making a decision ({\it i.e.} choosing C, D or S)   for each player depends on the ratio of the numbers of C, D and S bits stored in its memory.
Finally, all players update their memories simultaneously and the above process the system evolves by repeating the above procedure. Schematically, the description of the model is the following:
\begin{itemize}
\item Individuals are organized in the network defining the neighborhood of each player.

\item Each individual can choose among three strategies: C (collaborative), D (defective) or S (solitary).
We denote by $f_i$ the fraction of the $i$-type players in the network. The payoff matrix for any pair-game is given by

\begin{eqnarray}
  \mathbf{P} =
    \begin{blockarray}{cccc}
        &  \mathbf{C} & \mathbf{D} & \mathbf{S}  \\
      \begin{block}{c(ccc)}
        \mathbf{C}  &  1 & 1-r & q  \\
        \mathbf{D}  &  1+r & 0 & q \\
        \mathbf{S}  &  (1+ \xi)q & (1+ \xi)q & (1+ \xi)q \\
      \end{block}
    \end{blockarray}.
    \label{payoff}
\end{eqnarray}

The parameter $\xi$ is related to the fraction of S-players in the community at a given time, {\it i.e.}
$\xi = 1- f_S$. Both $r$ and $q$ are in the interval $[0,1]$. 
\item Each individual is endowed of a fixed-length memory of possible C, D or S strategies.
The initial memory is random  for all the players.

\item At any step, each individual plays with all its neighbors, the total gain being the sum of the pair-games.

\item The evolution of the system is memory-based. The players adopt a selfish attitude to update their strategies.
At any iteration, a player chooses its behavior selecting randomly one of the possible entries in its memory.
Then, it compares its total payoff with the possible gains coming from selecting the other two strategies. The strategy resulting in the highest gain is added to the player's memory,
and the oldest entry is deleted since the length is fixed. If two (or three)
strategies give the same (best) score, the update is selected randomly among the best choices.
\item The evolution then restarts with the next turn.
\end{itemize}

All simulations have been performed on regular lattices or random graphs with $10^4$ individuals. We report on numerical results where we kept the players' memory length $M=4$\footnote{The variation of the memory length affects the relaxation time of the system (see the last section), but has otherwise marginal effects on the strategic choices of the players in the steady state.}.
For the first and second phases, the  typical relaxation time of the system is of order of $T=1000$ iterations of the Monte Carlo simulations and the final fractions are always computed averaging from $T=1500$ to $T=2000$.

\section{Deepening the second phase dynamics}
\begin{figure}[htb]
  \centering
    \includegraphics[width=\columnwidth]{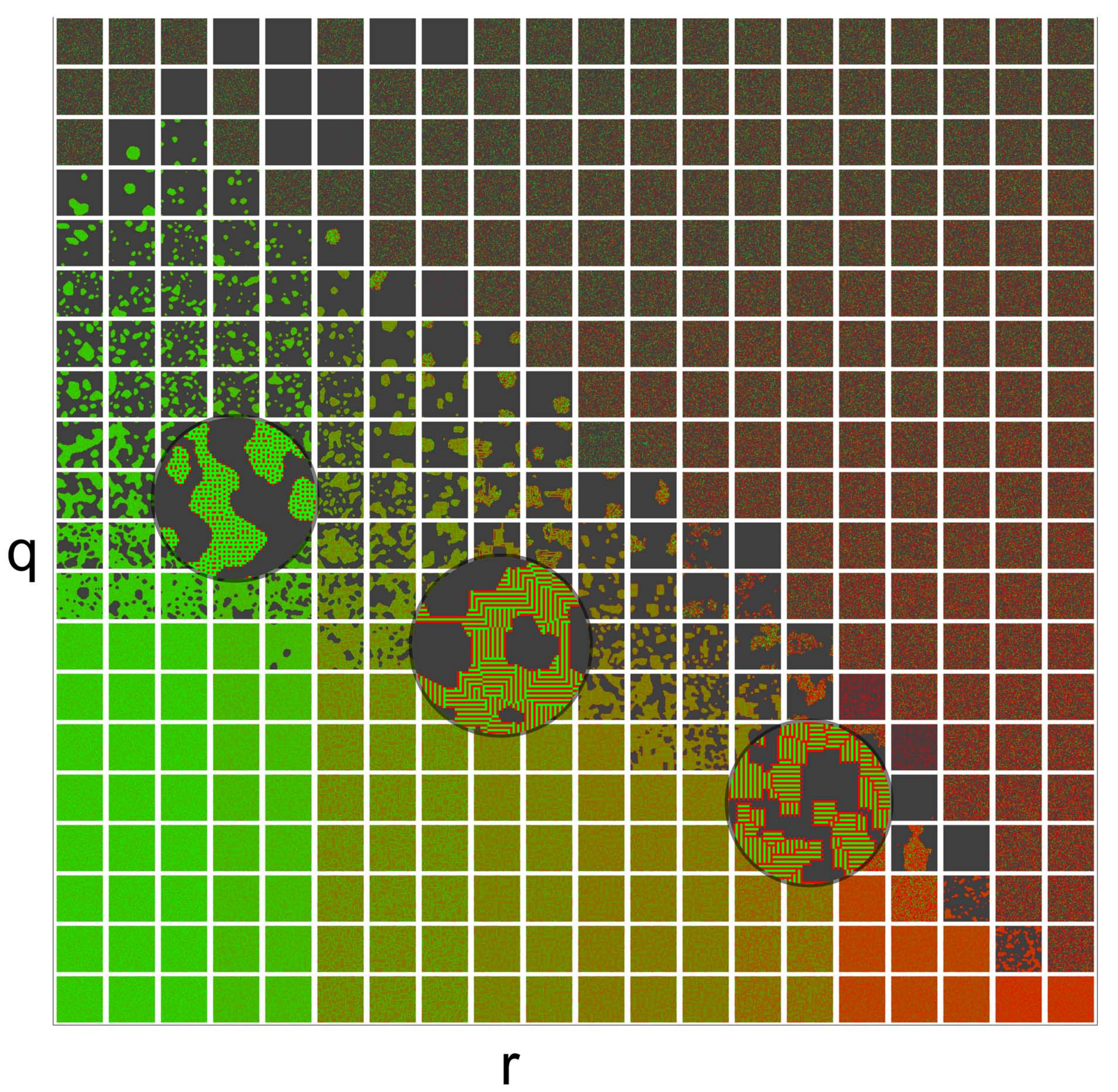}
      \caption{\label{fig3c} Collection of snapshots of final states of the game arranged in the $(r,q)$ plane. Collaborative players are
      denoted in green, defectors in red and loners in grey. Each snapshot illustrates a $100 \times 100$ lattice.
  }
\end{figure}
While the S-players gain a significant fraction of the total population as $r$ or $q$ grow (see figures $2$ and $3$ in the main text), their presence
has a relative effect on the local interaction among the C and D players. In figure \ref{fig3c} we show a collection of typical snapshots of final states
arranged in the $(r,q)$ plane. We notice the cluster arrangement of the loners and that
the regular patterns of C and D are preserved and persist also in the second phase (even when these solitary players are by far the dominant fraction).
The second phase comprises a number of subregions delimited by discontinuous jumps of the fractions,
depending on the details of the local interactions among the characters.
\begin{figure}[htb]
  \centering
    \includegraphics[width=\columnwidth]{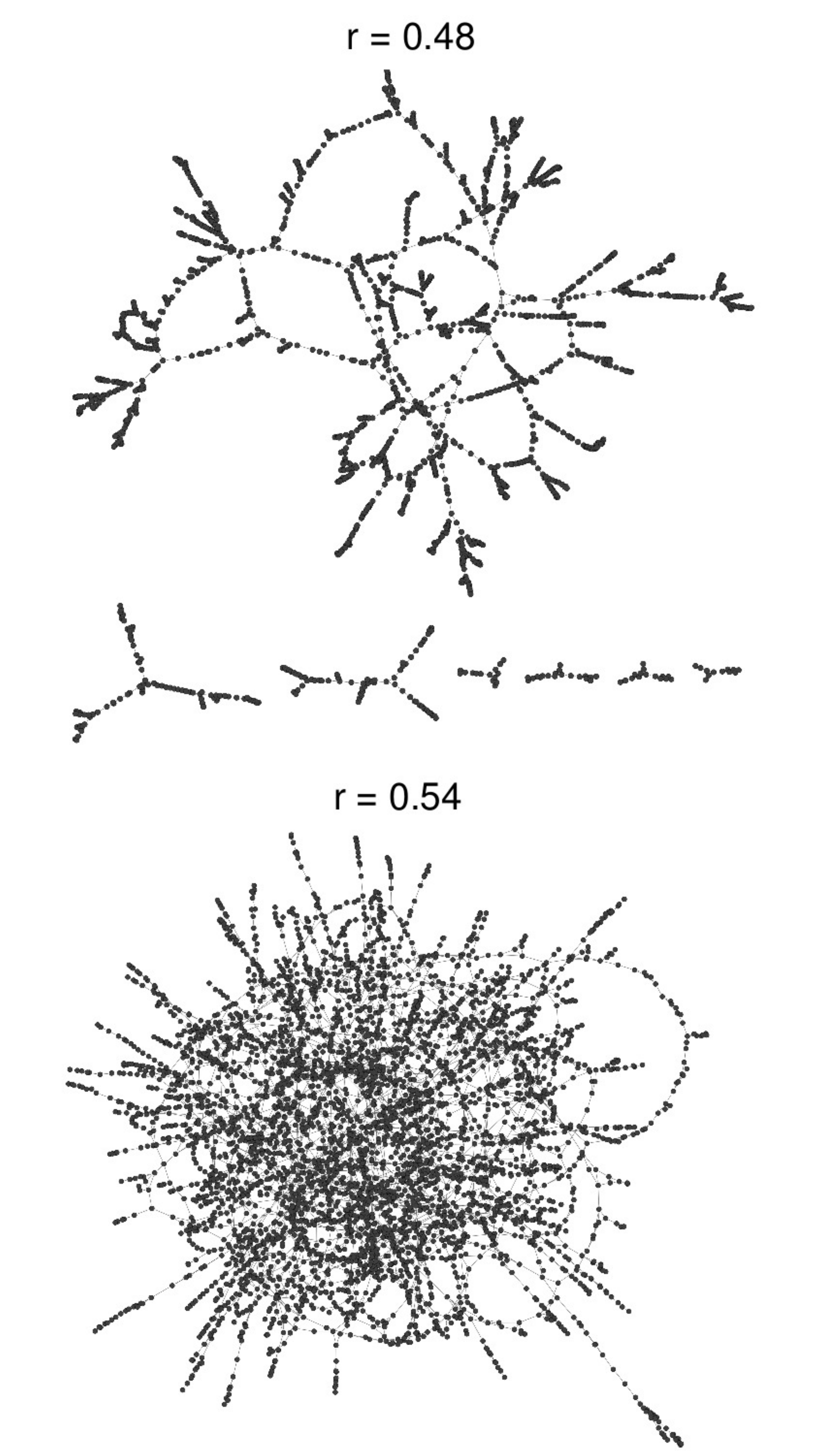}
      \caption{\label{SM3} Two sample of final configurations in Watts-Strogatz setting (with size $N=1000$ and rewiring probability $\theta=0.1$) with C- and D-nodes (and the relative links) removed. The tunable parameters are $q=0.3$ and $r=0.48,0.54$. These snapshots clearly show the strong tendency of the loners to clusterize in solitary islands (which eventually merge) in the second phase of the system dynamics.
  }
\end{figure}
Another important similarity with  the regular lattice is that S-players still exhibit a strong tendency to group together. An easy way to visualize this behavior is to look at the final configurations of the system (in the second phase), and remove from the graph all the C- and D-nodes (together with their links). Samples of the residual graphs are shown in figure \ref{SM3}, for $q=0.3$ and two choices of $r$ in the Watts-Strogatz setting. After the removal of the C- and D-players, the S-individuals never appear isolated, forming again clustered communities. In this respect, the small-world property of the network clearly favors the merging process of the {\em solitary islands}, since the size of the biggest connected component of S-players grows much more quickly than in the regular lattice. Behavioral similarities between mean evolution of order parameters within small worlds and regular lattices have been already highlighted also in the theoretical framework of classical statistical mechanics of spin systems \cite{Alain,SW-noi1,SW-noi2}.

Deepening the character of phase transitions, we tentatively termed them (respectively) first and second order since they appear in a quite standard way. Indeed, they show a strong resemblance of first and second order transitions in the Ehrenfest classification in equilibrium statistical mechanics. However, we stress that here there is no underlying {\em detailed balance} in players' dynamics, thus there is no guarantee regarding a convergence to a Gibbs measure for the persistence of the classical picture: this motivates our {\em mild} identification of these transitions.

\begin{figure}[htb]
  \centering
    \includegraphics[width=\columnwidth]{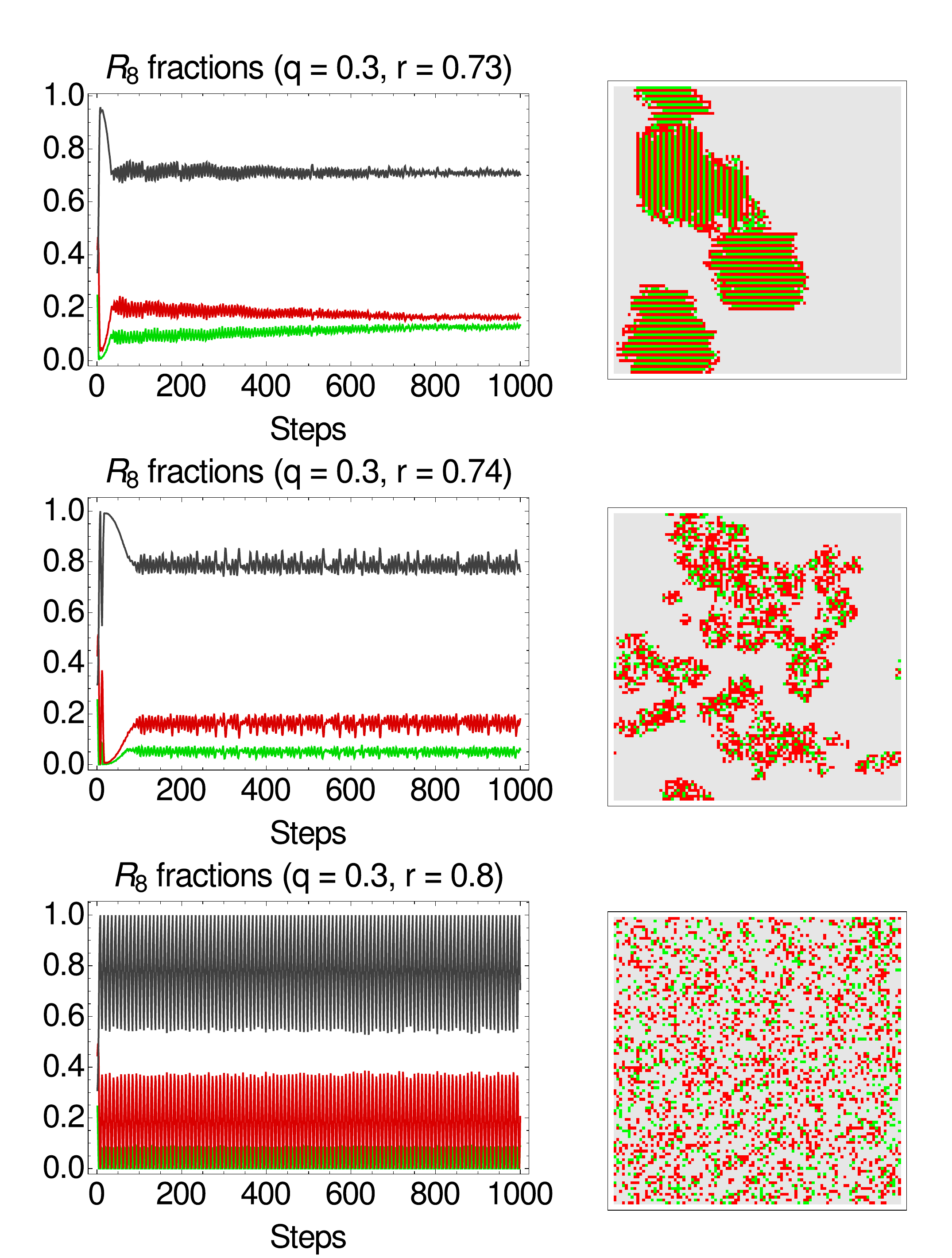}
      \caption{\label{figpermail} In the first column, we show the time evolution of the fractions of loners coop and defectors for fixed $q = 0.3$ and increasing $r$ towards the unstable region ($r = 0.73, 0.74, 0.8$). In the second column, we show three snapshots of configuration of the system. For $r = 0.73$, the system reaches an almost spatially stable configuration, with some residual oscillatory behavior of the players along the borders of the C/D regions. While the C/D player are confined in islands and the configuration is dominated by the solitary individuals, their typical spatial interaction pattern is still clearly recognizable. The time evolution of the fractions suggest that the oscillations are slowly decreasing in time, leading to a final stable configuration for very long times. Increasing the payoff to $r=0.74$, the oscillatory behavior becomes endemic. While it is still possible to clearly recognize patterns of C- and D-players in the configuration, the stability of their spatial interaction is lost. Finally, for even higher values of $r$ (and higher density of S-players), any hint of spatial arrangement is lost, and all the players behave erratically. Nevertheless, the fraction (widely) oscillates around a stable mean value for relatively long time averages.}
\end{figure}

\section{Deepening the third phase dynamics}

\begin{figure}[htb]
  \centering
    \includegraphics[width=\columnwidth]{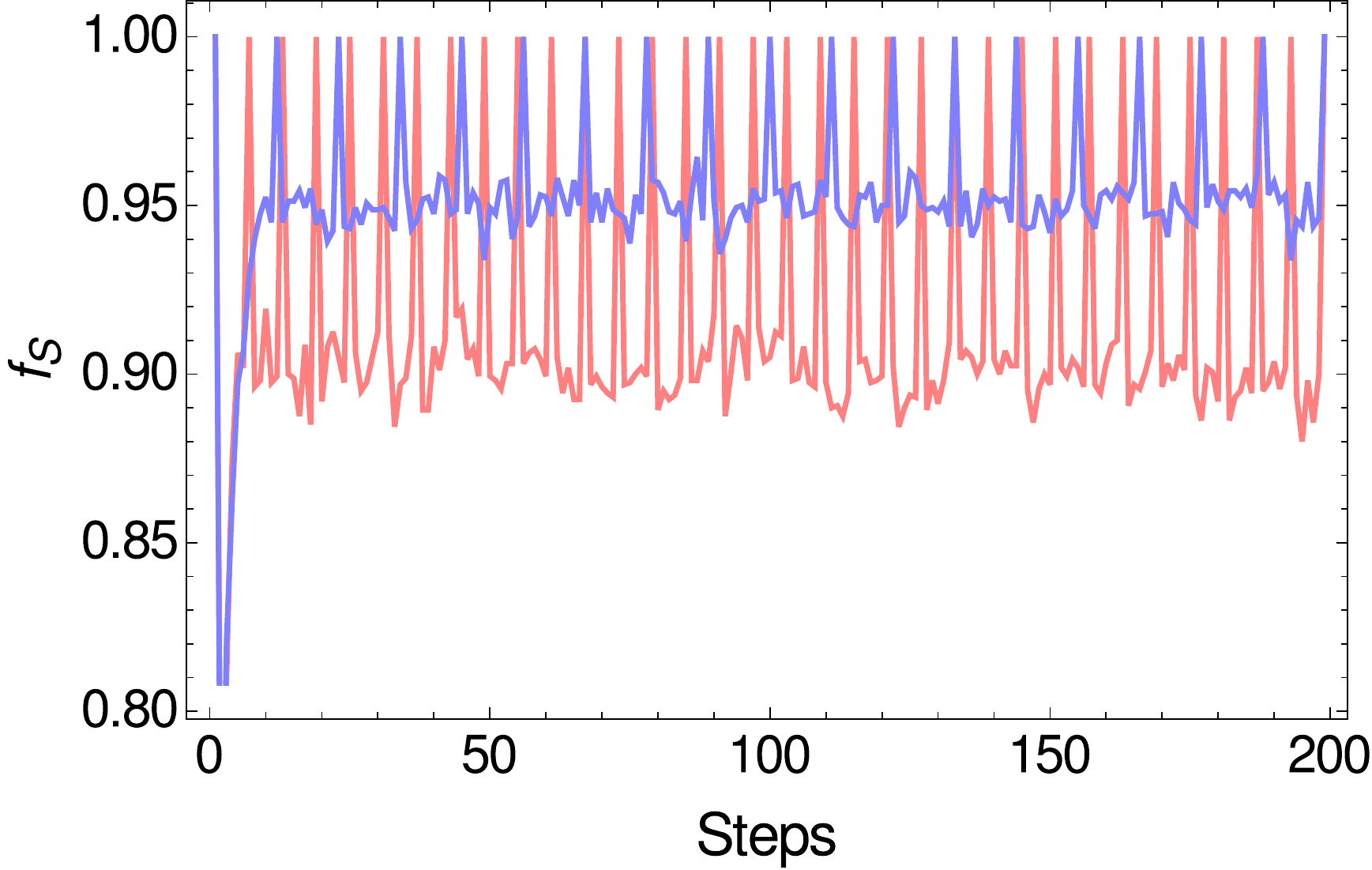}
      \caption{\label{quasiperiodo} Fractions of solitary players versus time for two different memory lengths ($M=5$ and $M=10$) in a simplified model with only C- and S-players for $q=0.87$. A trivial eye inspection reveals that the peak of maximal density are periodic, with typical frequency $M+1$.}
\end{figure}

While in the main text we have shown the intrinsic instability of the third phase (explaining its genesis and its relation to the global awareness feedback introduced by our extended payoff matrix in the limit of large $r$), here we aim to deepen the fact that - albeit the system is no longer stable - its three order parameters set on mean stable values if suitably averaged in time (see the high $r$ limit of all the plots in the figures $2,5,6$ in the main text).\footnote{We further stress that also in \cite{medie} the Authors average on suitably time-window the evolution of these fraction and they obtain analogous apparently stabilized averages.} At the basis of this successful processing of temporal averaging, there is the observation that (apart for stochastic effects), the network's dynamics at the onset of the third phase is rather {\em periodic}: there are waves of colonization by S-players that, once they reach their maximal amplitude, disintegrate and recursively give rise to novel generations. This process is governed by a {\em typical frequency} which can be clearly visualized in figures \ref{figpermail} and \ref{quasiperiodo}.
This typical frequency depends on the memory-length, since the distance between two peaks is given exactly by $M+1$ Monte Carlo steps. The explanation of its genesis is trivial: let us focus on $M=4$ with no loss of generality.  After an initial transient, for $q$ large enough the system evolves toward a final state made of by all S-players. Once this configuration is reached, all the players update their memory at random ({\it i.e.}, using the label $R$ to mean {\em randomly}, $\{S,S,S,R\}$). In successive steps, it is mandatory for all the players to keep choosing the $S$-strategy, so that their memory dynamics is trivially
\begin{eqnarray}
\{S,S,S,R\} &\to& \{S,S,R,S\}  \to  \nonumber  \\
&\to& \{S,R,S,S\} \to \{R,S,S,S\} \to \{S,S,S,S\} , \nonumber
\end{eqnarray}
up to the catastrophe again and so on: these dynamical regularities allow the effective thermalization to constant values for the fractions of players also in the third phase of the system.

\section{The two-component limit: solitary vs cooperative players}

\begin{figure}[htb]
	\centering
	\includegraphics[width=\columnwidth]{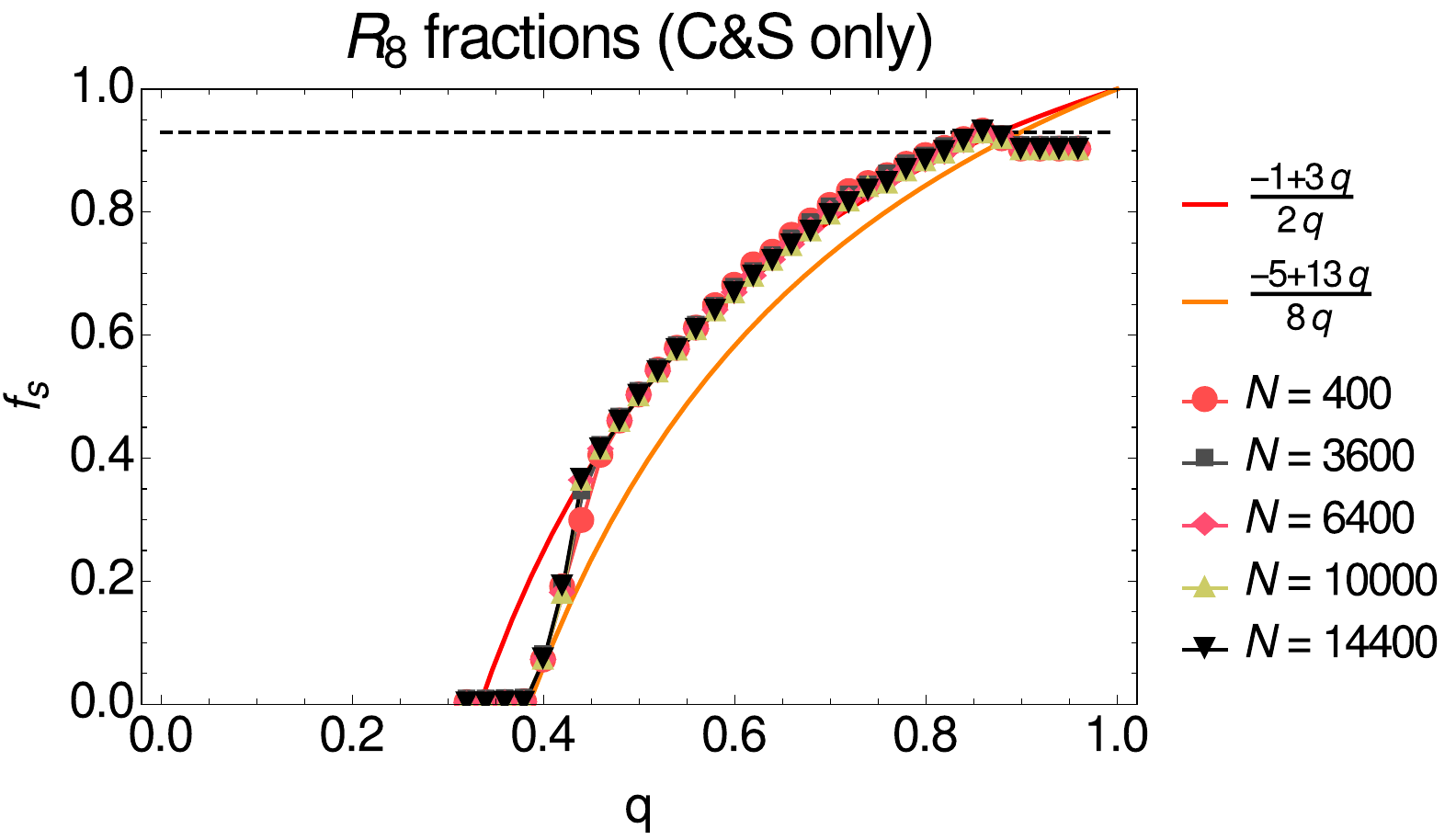}\\
	\caption{\label{reversed1}  By analyzing the mean growth of $f_S$ we conclude that, at fixed $r$ and evolving $q$, the solitary players conquer the system qualitatively always in the same manner, poorly dependent by the other player's activities. The plot shows the comparison between numerical data for the simplified model with only collaborative and solitary players and the relative theoretical predictions coming from the stability criterions.}
\end{figure}

By keeping $r$ fixed and moving $q$, we can check the evolution of the three fractions of C-,\ D-,\  and S-players. The S-density grows smoothly  and its growth is  largely determined by the domain-walls lying on the boundaries of their clusters (and by the advantages a C- or D-player has in turning into an S-player on that border).
\newline
To deepen this feature, we restricted the game to a two-players model, with solely C- and S-players and with a payoff matrix given by
\be\small
  \mathbf{P}_{SG} =
    \begin{blockarray}{ccc}
        &  \mathbf{C} & \mathbf{S}  \\
      \begin{block}{c(cc)}
        \mathbf{C}  &  1 & q  \\
        \mathbf{S}  &  (1 + \xi) q & (1 + \xi) q   \\
      \end{block}
    \end{blockarray}.
\ee
In a nutshell, this setup trivializes the interactions among C- and D-individuals effectively resulting in a simplified model for the dynamics of the S-players, whose density can finally be plotted versus $q$ as shown in figure \ref{reversed1}. The critical line is indeed achieved by exploiting the stability criterion
\begin{equation}
8 (1 +(1-f_S)) q > 4 + 4 q,
\end{equation}
for a C-player lying on the boundaries of an S-cluster.
\newline
The red curve in figure \ref{reversed1} can be obtained by studying the behavior of a single player in a $3 \times 3$ triangle placed at the boundary of a solitary cluster and checking the threshold for its shift \mbox{S $\rightarrow$ C}. In fact, if the player keeps the S-strategy, its total payoff would be
\begin{equation}
SL = 8\left(1 + \left(1-f_S\right)\right) q
\end{equation}
while, if it adopts for the C-strategy, it would be
\begin{equation}
SC = 4 + 4 q,
\end{equation}
since it would have four C-peers and four L-peers. Solving the equality $SC = SL$ w.r.t. the solitaries fraction we obtain the critical S-density $f_S(q) = (-1 + 3 q)/(2 q)$, which is the equation for the red curve in figure \ref{reversed1}.
\newline
The orange curve can be similarly obtained with evaluating the threshold for a transition C $\rightarrow$ S near the boundary of an S-cluster. With the same reasoning as before, we get the following criterion:
\begin{equation}
8 (1 + (1- f_S)) q = 5 + 3 q,
\end{equation}
and solving again w.r.t $f_S$, we get the equation
\begin{equation}
f_S (q)=\frac{-5+13q}{8q}.
\end{equation}

\begin{figure}[htb!]
	\centering
	\includegraphics[width=1.0
	\columnwidth]{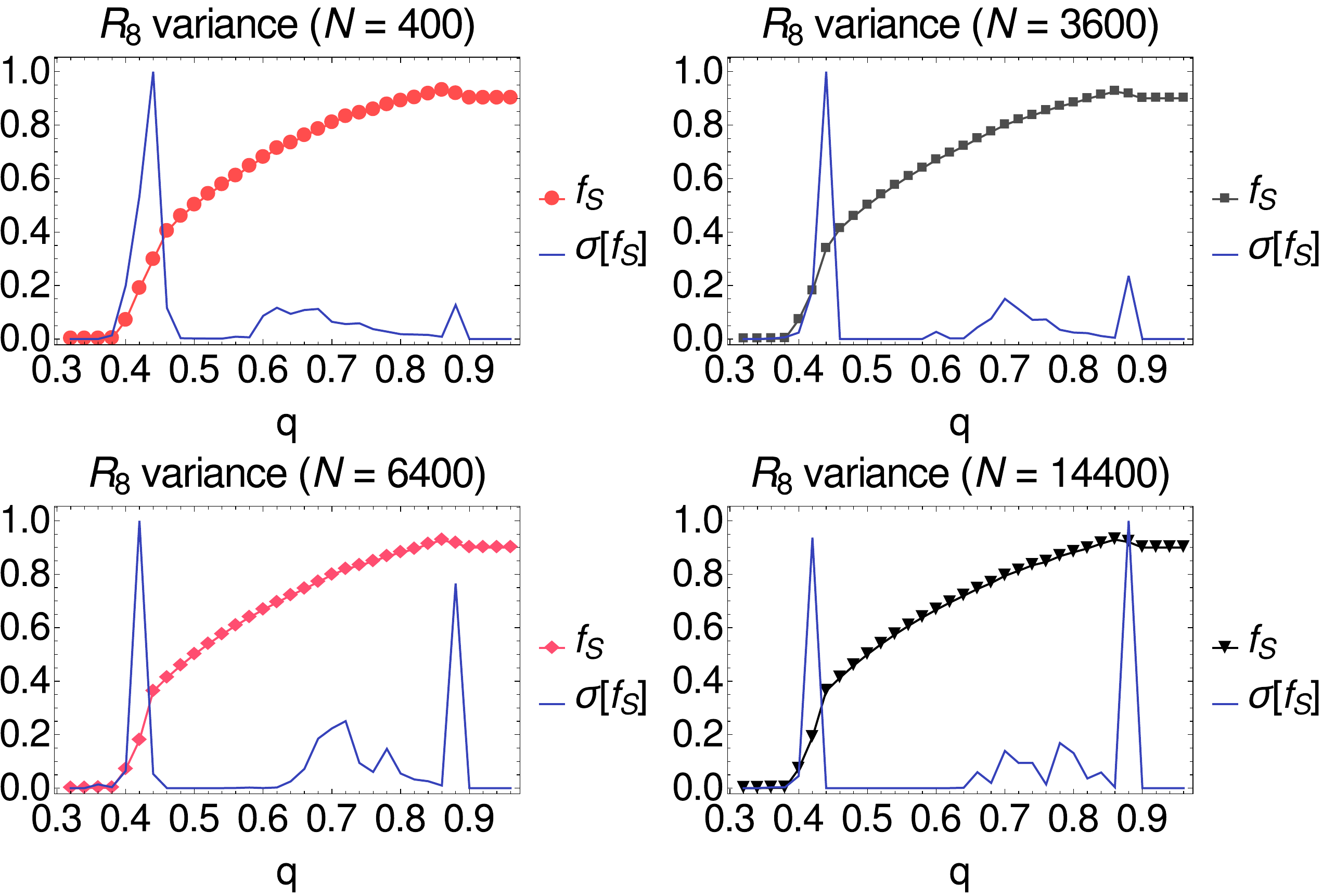}\\
	\caption{\label{varianze}  In these four panels we show, for four different network sizes, how the evolution of the mean of the order parameter $f_S$ is coupled with the growth of its standard deviation: as it shines trough all the plots, the raise from zero of $f_S$ is smooth and it is always accompanied by a peak in its variance, (i.e. the normalized {\em solitary susceptibility}). We note further that - at the onset of the unstable phase - again the  fluctuations of $f_S$ peak (more mildly than in the first transition, yet, somehow they give a critical flavour also to that transition). Note that, among these two main peaks, a plethora of small fluctuations further characterize several local critical rearrangements.}
\end{figure}
\begin{figure}[htb!]
	\centering
	\includegraphics[width=1.0
	\columnwidth]{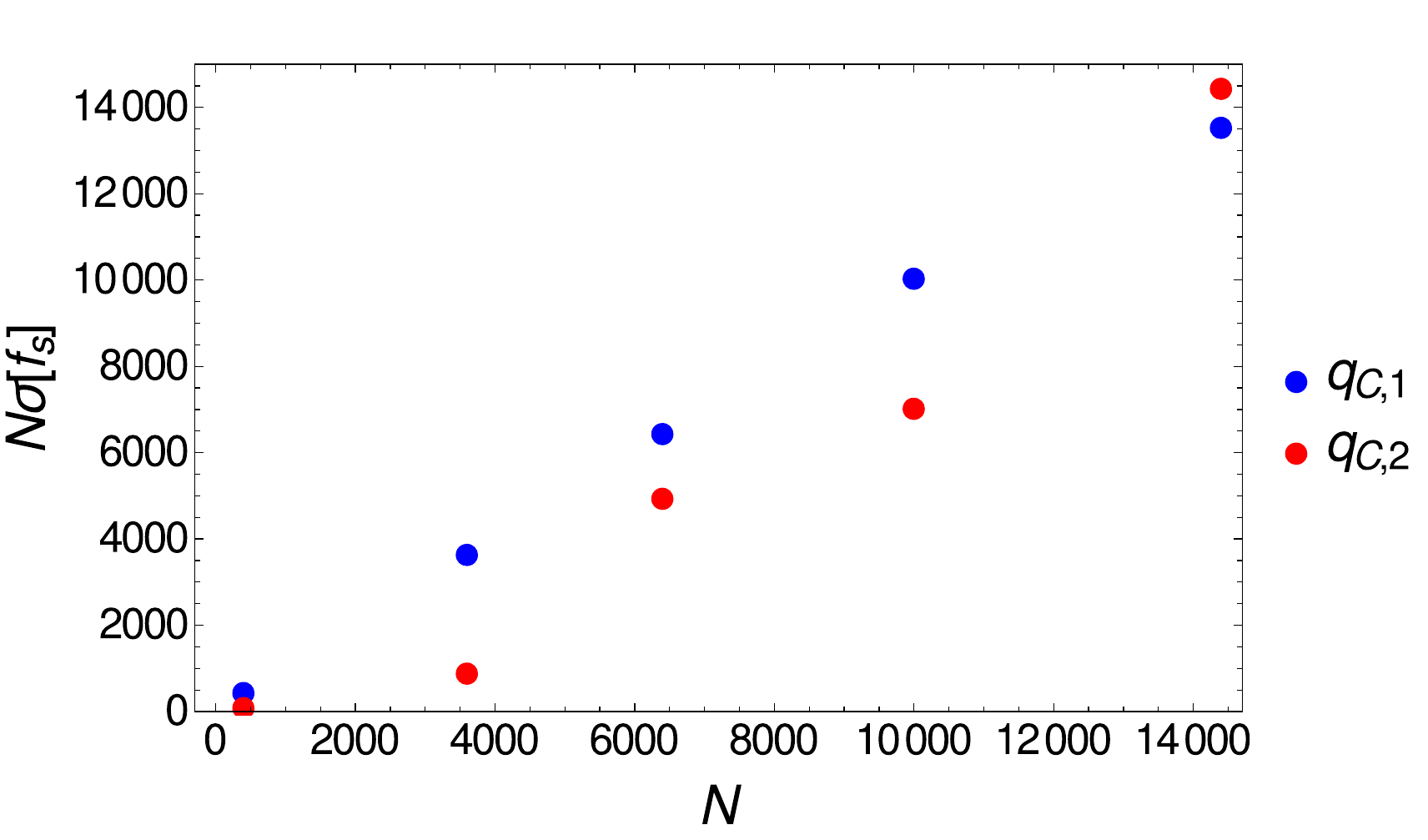}\\
	\caption{ \label{FSS} The plot shows the dependence of the solitary susceptibility ({\it i.e.} the variance of the $S$-density suitably amplified with a volume factor) as a function of the network size $N$ at the critical values $q_{C,i}$ in the simplified model with only collaborative and solitary players. Here, $q_{C,1}$ is the critical value for the transition between the first and second phase of the dynamics, while $q_{C,2}$ is the critical value for the transition in the unstable third phase. By looking at its scaling behavior at these critical values, it is reasonable to expect that the solitary susceptibility diverges in the infinite size limit, meaning that the transitions in $q$ can be classified as second order phase transition.}
\end{figure}
Finally, in this simplified model we also studied the character of the phase transitions in $q$, finding that they can be identified as second order in the standard Ehrenfest classification, see figures \ref{varianze} and \ref{FSS}.

\section{Memory relaxation of single player games}

In order to analyze the dependence on $M$ of the relaxation time of the memory-based game, let us
discuss the simple case of a single player self-game. The state $s$ of the player is a vector with $M$ bits taking 0 or 1 values. At each step, we extract a move $\mu\in\{0,1\}$ with probability $n_{\mu}(s)$ proportional
to the number of $\mu$ values in $s$.
Then, the memory is updated $s\to s'$, where $s'$ is obtained from $s$
by first shifting and then adding the new value $\overline \mu$, where $\overline 0 = 1$ and $\overline 1 = 0$.
This is a finite Markov chain with dimension $\text{d}=2^{M}$. By analizing the transfer matrix
$T_{s', s}$, it can be shown that it has only one stable eigenvector with eigenvalue $\lambda_{0}=1$. This is a stationary non-trivial
probability distribution $P_{s}$ given by
\be
P_{s}= \frac{\sqrt\pi}{4^{M}}\,\frac{\Gamma(M+1)}{\Gamma(M+\frac{1}{2})}\binom{M}{n_{1}(s)}.
\ee
The characteristic polynomial of $T$ factorizes and the eigenvalue $\lambda_{1}$ with largest modulus (
strictly smaller than 1) is the largest modulus solution of the polynomial equation
\be
\label{A.2}
M\,\lambda_{1}^{M}+\frac{\lambda_{1}^{M}-1}{\lambda_{1}-1}=0.
\ee
At large $M$, this gives
\be
\label{A.3}
|\lambda_{1}| = 1-\frac{\alpha}{M}+\cdots,
\ee
where $\alpha = 1.58255\dots$ is obtained from $\alpha=\log(-\beta\csc\beta)$ with
$\beta$ being the root of
\be
-1+\log(-\beta\,\csc\beta)+\beta\cot\beta=0,
\ee
 in $[\pi, 2\pi]$. This means that the game relaxed to the equilibrium distribution with a characteristic number of
iterations $N_{\rm relax} = -1/\log|\lambda_{1}| = M/\alpha+\cdots$. A model closer to our situation is characterized by the storage of $\mu$ (instead of $\overline \mu$). In this case there are two
stable eigenvectors associated with the stable configurations with all 0's or all 1's. The subleading
eigenvalue $\lambda_{1}$ cannot be computed in closed form parametrically in $M$, as in (\ref{A.2}),
but can be numerically solved at finite $M$ with results similar to (\ref{A.3}) (although with a different
coefficient $\alpha$).

%
%
%
%

\end{document}